\documentclass[superscriptaddress,showpacs,preprintnumbers,amsmath,amssymb,
nofootinbib,twocolumn,aps,prd,10pt]{revtex4-1}
\usepackage{amssymb,amsmath}
\usepackage{epsfig}
\usepackage{xcolor}
\usepackage[utf8]{inputenc}
\usepackage{stmaryrd}
\usepackage{mathrsfs}
\usepackage{mathalfa}
\usepackage[normalem]{ulem}

\begin{document}

\title{Dynamical Origin of the Entropy of a Black Hole}

\author{Valery P. Frolov}
\email{vfrolov@ualberta.ca}
\affiliation{Theoretical Physics Institute, University of Alberta, Edmonton, Alberta, Canada T6G 2E1}
\author{Igor Novikov}
\email{novikov@nordita.dk}
\affiliation{The Copenhagen University Observatory,
\O ster Voldgade 3, DK-1350 Copenhagen K, Denmark,
Nordita, Blegdamsvej 17, DK-2100 Copenhagen \O , Denmark, and
Astro Space Centre of P.N. Lebedev Physical Institute,
Profsoyuznaya 84/32, Moscow, 117810, Russia}



\begin{abstract}
Modes of physical fields which are located inside a horizon and  which
cannot be observed by a distant observer are identified with dynamical
degrees  of  freedom  of  a  black  hole.  A new invariant statistical
mechanical definition  of a  black-hole's  entropy  is  proposed.   It
is   shown   that   the main  contribution to  the entropy is given by
thermally excited  `invisible' modes propagating in the close vicinity
of  the  horizon.  A   calculation  based  on  the proposed definition
yields a value of the  entropy which is   in   good  agreement    with
the   usually  adopted    value $A^H /(4l^2_{\mbox{\scriptsize{Pl}}})$,
where  $A^H$ is  a  black  hole  surface  area  and
$l_{\mbox{\scriptsize{Pl}}}$ is the Planck's length.
\end{abstract}

\maketitle
\newpage

According to the  thermodynamical analogy in  black hole physics,  the
entropy of a black hole  is defined as
$S^H =A^H  /(4l_{\mbox{\scriptsize{Pl}}}^2)$,
where   $A^H$   is   the   area   of   a   black   hole   surface  and
$l_{\mbox{\scriptsize{Pl}}}=(\hbar G/c^3~)^{1/2}$ is the Planck's length
\cite{Beke:72,Beke:73}. Hawking's theoretical discovery \cite{Hawk:75}
of quantum black hole  evaporation proved the  reality of the  black hole
temperature and   fixed the  coefficient relating  entropy of  a black
hole with  its surface  area.   The generalized  second law  (i.e. the
statement that the sum $S=S^H +S^m$ of a black hole's entropy and  the
entropy $S^m$ of the outside matter cannot decrease) implies that  the
black hole's  entropy plays  the same  role as  the usual  entropy and
shows up to which extend the  energy contained in a black hole  can be
used to produce work  \cite{Beke:72,Beke:73,Beke:74}. Four laws of black
hole physics which form the basis in the thermodynamical analogy  were
formulated in \cite{BaCaHa:73}.

Despite some promising attempts
\cite{Beke:73,Beke:80,York:83,ZuTh:85,ThPrMa:86}, the dynamical
(statistical mechanical) origin of a black hole entropy has not been
well understood.  Bekenstein \cite{Beke:73} who introduced the notion of
the  black  hole  entropy  related   it  with  `the  measure  of   the
inaccessibility of information (to  an exterior observer) as  to which
particular  internal  configuration  of  the  black  hole  is actually
realized in  a given  state' (i.e.   for given  values of  black  hole
parameters: mass,  charge, and  angular momentum.)   According to  the
`standard' interpretation  these different  internal states  of a
black hole are related with different possible initial conditions
which  may result  in  the  creation  of  a  stationary  black hole
with the same parameters \cite{Beke:73}.  In this  approach the
entropy    of a black   hole   is  considered     as  the  logarithm
of the number of distinct   ways    that    the   hole    might
have   been     made \cite{ZuTh:85,ThPrMa:86}.

This  definition  of  the  black-hole  entropy  resembles up to some
extend the usual definition of the  entropy of matter. But there is  a
big  difference  which  makes  the above `standard' interpretation not
completely  satisfactory.   The entropy of  matter is connected   with
its real internal dynamical degrees  of  freedom  which exist at given
moment and which can be affected by an external force. By  getting
information   about  the  states  of  some of the
internal degrees of freedom one can  reduce the  entropy  of   matter.
(The   total   entropy   of   matter   and     its  environment never
decreases.) The loophole of the above described  `standard'
interpretation is that  one cannot  indicate dynamical   degrees
of freedom of a black hole which are responsible for its entropy.
Even if an   observer is
falling  down into  a  black  hole  long  time  after  its   formation
he cannot receive   more information   about the  initial state  than the
exterior observer. The  information concerning the  initial conditions
is lost for the interior observer for the same  reason as it is   lost
for  an  exterior   observer  \cite{DoNo:78}. The collapsing body
and its structure become invisible for the late-time observer and
only a few macroscopic parameters (mass, angular momentum,
and charge) remains measurable. This situation
reminds the famous grin of the Cheshire Cat remaining after
the Cat himself had disappeared. In  this  sense different
possible  initial  states  for  a  black  hole  are  `Cheshire-Cat's'
variables \cite{note1}.

This difficulty  of the  `standard' interpretation  becomes especially
vivid if one considers a recently  proposed  gedanken  experiment   in
which  a  traversable  wormhole  is  used   to  get   information from
black-hole's interior   \cite{FrNo:93}.  Namely  it was shown  that the
area  of  the surface   of a   black hole   decreases if   one of  the
mouths  of  a traversable wormhole   is falling  into   a black  hole
while the other mouth remains outside it.
This decrease  continues until the  other mouth   crosses the
horizon or the wormhole is destroyed.  After   this the  surface area  of
a black hole  returns back  to its initial value.  There is an  evident
mystery in such a behavior  if in accordance with the   usual rules
one identifies the  entropy of a   black hole with the loss of information
about the possible origin of a black hole, because the wormhole does
not get any new information about it.     One  can escape  a
contradiction  with   the generalized  second  law only  if   one
assumes that there exist some real internal degrees of freedom
hidden inside the black hole so that   additional new information
about     the internal states   of  a black hole can be obtained
by using a wormhole during this process.

York \cite{York:83} tried to  solve the problem of  `Cheshire-Cat's'
variables by proposing that  quasi-normal modes of a black  hole
can play the role  of its dynamical degrees  of freedom. This attempt
also cannot be considered  as completely satisfactory.  The entropy of
the quasi-normal modes which are thermally excited at the given moment
of time is much  smaller than  $A^H/(4l_{\mbox{\scriptsize{Pl}}}^2)$ (the
adopted
value for the  entropy of a  black hole). In  order
to obtain  the   black  hole's    entropy  York \cite{York:83}  makes  an
additional assumption that the entropy ``results  when we  add up  all
the excited (normal modes) states that  disappear if   we allow the
hole  to  evaporate down to  a final mass  zero''. In other words the
black  hole's entropy is  again  related  not  with  dynamical  degrees
of freedom which are really  excited  at  the  given  moment  of  time
but with a number of different possibilities to excite them during the
life-time of a black hole.

In this paper we propose a new approach to the problem of  statistical
mechanical calculation of the entropy  of a black hole. The  main idea
of this approach is to identify the dynamical degrees of freedom of  a
black  hole  with  those  states  of all physical (quantum) fields which
are located  inside   a black   hole   and   cannot   be   seen  by  a
distant observer.   An   excitation    of   these   states   does  not
change   the    external parameters  of the    black  hole. For  fixed
external    (macroscopic)  parameters      there      exist       many
microscopically      different   (internal)     states           which
cannot       be   distinguished    by observations made in the
exterior region.   Due  to  quantum
effects  these   internal  states become  thermally excited  and  give
contribution to  the black hole  entropy.  This makes  the  definition
of the   entropy and   other thermodynamical characteristics  of black
holes quite similar to those adopted in
the usual  statistical  mechanical description
of matter. In the absence of mutual interactions one can consider  the
contribution  to  the  entropy  of  each of physical fields (including
gravitational  perturbations)  independently.  It  will be shown that
the  resulting   entropy of  a black  hole does  not depend on the
number of fields.

In order to make the definition of the black hole entropy more concrete
we assume that there exists a stationary black hole and denote by
$\hat{\rho}^{\mbox{\scriptsize{init}}}$ the density matrix describing in
the Heisenberg representation  the initial state  of  quantum  fields
propagating in its background.  One may consider e.g. the in-vacuum
state for a black hole evaporating in the vacuum, or the Hartle-Hawking
state for a black hole in equilibrium  with  thermal  radiation.   For
an exterior observer the system under  consideration consists  of two
parts: a  black hole and radiation outside of it. The state of
radiation outside the black hole is  described   by  the   density
matrix   which  is   obtained  from $\hat{\rho}^{\mbox{\scriptsize{init}}}$ by
averaging it over the states which are located inside the black hole
and are invisible in its exterior
\begin{equation}
\hat{\rho}^{\mbox{\scriptsize{rad}}}
=\mbox{Tr} ^{\mbox{\scriptsize{inv}}}\hat{\rho}^{\mbox{\scriptsize{init}}}.
\label{1}
\end{equation}
For an isolated black hole this density matrix
$\hat{\rho}^{\mbox{\scriptsize{rad}}}$
in particular describes its Hawking radiation at infinity.

Analogously we define the density matrix describing the state of a
black hole as
\begin{equation}
\hat{\rho}^H =\mbox{Tr}
^{\mbox{\scriptsize{vis}}}\hat{\rho}^{\mbox{\scriptsize{init}}}.  \label{2}
\end{equation}
The trace-operators $\mbox{Tr} ^{\mbox{\scriptsize{vis}}}$ and
$\mbox{Tr} ^{\mbox{\scriptsize{inv}}}$ in these
relations mean that the trace  is taken over the states located either
outside  (`visible')  or  inside   (`invisible')  the  event   horizon
correspondingly.  We define the entropy of a black hole as
\begin{equation}
S^H =-\mbox{Tr} ^{\mbox{\scriptsize{inv}}}(\hat{\rho}^H \ln \hat{\rho}^H ).
\label{3}
\end{equation}
The proposed definition of  the entropy of  a black hole  is invariant
in   the  following   sense.    Independent   changes   of      vacuum
definitions  for `visible' and  `invisible' states  do not  change the
value   of   $S^H$.    Bogolubov's   transformations    describing  an
independent changes  of the vacuum states inside and outside  the  black
hole  can  be   represented  by  the         unitary          operator
$\hat{U}=\hat{U}^{\mbox{\scriptsize{vis}}}
\otimes \hat{U}^{\mbox{\scriptsize{inv}}}$         where
$\hat{U}^{\mbox{\scriptsize{vis}}}$         and
$\hat{U}^{\mbox{\scriptsize{inv}}}$ are   unitary
operators   in  the  Hilbert  spaces    of `visible'   and  `invisible'
particles,   correspondingly.    The above  used trace  operators  are
invariant under such transformations.

In order to define the states one usually use modes expansion.  The modes are
characterized by a complete set of quantum numbers. Due to the symmetry
properties  one can choose such a subset $J$ of quantum numbers
connected with conservation laws (such as orbital and azimuthal
angular momenta, helicity and so on) that
guarantees the factorization of the density matrices.
In the absence of mutual interaction of
different fields the subset $J$ necessarily includes  also the parameters
identifying the type of the field (e.g. mass, spin, and charge).
The factorization in particular means that
\begin{equation}
\hat{\rho}^{\mbox{\scriptsize{init}}}=\otimes _J
\hat{\rho}^{\mbox{\scriptsize{init}}}_J ,\label{4}
\end{equation}
where $\hat{\rho}^{\mbox{\scriptsize{init}}}_J$
is acting in the Hilbert space ${\cal H}_J$ of states with the chosen
quantum numbers $J$, while the complete Hilbert space is ${\cal
H}=\otimes _J {\cal H}_J$. The factorization also means that the
separation into `visible' and `invisible' states can be done
independently in each subset of modes with a fixed $J$ so that
\begin{equation}
\hat{\rho}^H =\otimes _J \hat{\rho}^H_J \hspace{1cm} \hat{\rho}^H_J
=\mbox{Tr}^{\mbox{\scriptsize{vis}}}_J
\hat{\rho}^{\mbox{\scriptsize{init}}}_J,
\label{5}
\end{equation}
\begin{equation} S^H =\sum_J
S_J^H ,\hspace{1cm} S_J^H =-\mbox{Tr} ^{\mbox{\scriptsize{inv}}}_J
(\hat{\rho}^H_J \ln
\hat{\rho}^H_J ). \label{6}
\end{equation}
where all the operators with subscript $J$
are acting in the Hilbert space ${\cal H}_J$.

We begin by calculating $S^H$ for a non-rotating uncharged black hole.
We suppose that the black hole is contained inside a spherical cavity
$B$ of radius $r_0$ with mirror-like boundary. We choose $r_0$ small
enough to guarantee the stable equilibrium of a black hole with
thermal radiation inside the cavity.  Instead of physical
metric $ds_{\mbox{\scriptsize{phys}}}^2$ it is convenient to use its
dimensionless
form $ds^2 =r_g^{-2}ds_{\mbox{\scriptsize{phys}}}^2$ ,

where $r_g =2M$ and $M$ is
the mass of the black hole. In the Kruskal coordinates $(U,V)$ the
metric $ds^2$ reads
\[
ds^2 =-2BdUdV+x^2 d\omega ^2,\hspace{1cm}
d\omega ^2=d\theta ^2 +\sin ^2 \theta d\phi ^2 ,
\]
\begin{equation}
B=2x^{-1}\exp (1-x),\hspace{1cm}UV=(1-x)\exp (x -1), \label{7}
\end{equation}
where $x=r/r_g$. The Killing vector $\xi $ normalized to unit at infinity
in the metric $ds^2$ is
\begin{equation}
\xi ^{\mu}\partial _{\mu}=1/2 (V\partial _V -U\partial _U ). \label{8}
\end{equation}

For simplicity we consider  a conformal massless scalar field
$\hat{\varphi}$ obeying the equation
\begin{equation}
\Box \hat{\varphi} -1/6 R\hat{\varphi}
=0. \label{9}
\end{equation}
The following two sets $f^{UP}$ and $f^{SIDE}$ of classical complex
solutions of this equation which we call  $UP$ and $SIDE$ modes
\begin{equation}
f^{UP,SIDE}_{\nu lm}=x^{-1}F^{UP,SIDE}_{\nu
l}(U,V)Y_{lm}(\theta ,\phi ), \label{10}
\end{equation}
\begin{eqnarray}
&&\xi ^{\mu}\partial _{\mu}F^{UP}_{\nu l}=-i(\nu /2) F^{UP}_{\nu l},\nonumber\\
&& \xi ^{\mu}\partial _{\mu}F^{SIDE}_{\nu l}=i(\nu /2)
F^{SIDE}_{\nu l}, \nonumber
\end{eqnarray}
will be used in our consideration. The functions
$F^{UP}_{\nu l}$ and $F^{SIDE}_{\nu l}$ obey the equation
\begin{equation}
-\partial _U \partial _V F_{\nu l} -W_l F_{\nu l} =0, \label{11}
\end{equation}
\begin{equation}
W_l \equiv x^{-3} \exp (1-x) [l(l+1)+1/x]. \label{12}
\end{equation}

The $UP$ modes are radiated  in the exterior space  from
horizon  and vanish at the past null infinity $\cal{J}^-$,
while
the $SIDE$  modes are  radiated by the inner part ($U>0$) of
the horizon  into the black hole's
interior \cite{note2}.
We  denote  by  $\hat{\alpha}$  and  $\hat{\alpha}^*$
with  the corresponding superscripts the operators of annihilation and
creation of  particles  in  these   modes.   The  normalized
wave-packet-type solutions $f_J$ $(J=jnlm)$ are constructed from modes
$f_{\nu lm}$  as follows ($\delta >0$)
\begin{equation}
f_{jnlm}=\delta ^{-1/2}\int ^{(j+1)\delta}_{j\delta}
\exp (2\pi in\nu /\delta )f_{\nu lm}d\nu .\label{13}
\end{equation}

We denote by $\nu _J$ the average frequency of the wave packet $J$ and
introduce  two sets of new modes $f_J^B$ and $f_J^W$ by the
relations
\[
f^B_J =[1-w_J^2]^{-1/2} [f^{SIDE}_J +w_J \bar{f}^{UP}_J ],
\]
\begin{equation}
f^W_J =[1-w_J^2]^{-1/2} [f^{UP}_J +w_J \bar{f}^{SIDE}_J ], \label{14}
\end{equation}
where  $w_J=\exp (-\pi \nu_J )$.  These modes are of positive frequencies
with respect to the affine parameter along the horizon $H^-$.
In the presence of mirror-like boundary surrounding the black hole
the vacuum state with respect to $B$ and $W$ modes coincides with
the Hartle-Hawking state.
The density matrix corresponding to the Hartle-Hawking state takes the
form (\ref{4}) with
\begin{equation}
\hat{\rho}^{\mbox{\scriptsize{init}}}_J =\rho ^0_J
:\exp \left[ -\hat{\alpha}_J^{*B} \hat{\alpha}_J^B
-\hat{\alpha}_J ^{*W} \hat{\alpha}_J^W \right] : , \label{15}
\end{equation}
where $\rho ^0_J$ is the normalization constant, $:(\ldots ):$ means normal
ordering with respect to operators $\hat{\alpha}^*$ and $\hat{\alpha}$ which
enter the expression (\ref{15}). For the Hartle-Hawking state the $UP$-modes
(as well as $SIDE$-modes) are thermally excited. That is why for an eternal
black hole the density matrix (\ref{15}) describes the equilibrium state
of a black hole with thermal radiation in its exterior. If the black hole
is not eternal but is formed in the process of the gravitational collapse
the expression for $\hat{\rho}^{\mbox{\scriptsize{init}}}$ should be
modified. This modification is important for the modes emitted at the
time close to the moment of black-hole's formation.
For late-time modes the expression
(\ref{15}) always provides almost exact description .
That is why instead of calculating  the entropy of a stationary
black hole long after its formation it is possible to calculate
the entropy of an eternal black hole with the same parameters and for
the initial state decribed by the density matrix (\ref{15}). We use this
(technically more simple) approach.

Consider a null surface $V=v$. Its intersection with the horizon
represents the surface of the black hole at a
given moment of time. Denote by $\Sigma _v$ a part of this null surface
lying between the horizon  and the boundary of the cavity B.
It is evident that $SIDE$-modes  being propagated inside the black hole
never cross $\Sigma _v$ and
hence according to our definition they are `invisible'. Those
$UP$-modes which reach $\Sigma _v$ are to be considered as `visible'.

We introduce new coordinates $\eta, \xi$ in the  exterior  ($U<0$, $V>0$)
region, where  $UP$-modes are propagating
\begin{equation}
\eta =-\frac{1}{2}\ln (-V/U), \, 
\xi =\frac{1}{2}q(x),\, q(x)\equiv x-1 +\ln (x-1). \label{16}
\end{equation}
Then the wave equation (\ref{11}) for modes $F^{UP}_{\nu l}$ takes the
form
\begin{equation}
[-\partial _{\eta}^2 +\partial _{\xi}^2 -V_l] F^{UP}_{\nu l}=0, \label{17}
\end{equation}
where $V_l =4(x-1)x^{-3} [\mu _l^2 -1+x^{-1}]$,\hspace{.3cm}
$\mu_l^2 =l(l+1)+1$.
For $l\gg 1$ the potential $V_l$ has maximum $\approx 16 \mu _l^2 /27$
at $x=x_m \approx 3/2$.  We call `escaping' (or briefly E-modes) the
$UP$-modes with $\nu > 4\mu _l /\sqrt{27}$ which (in the absence of the
boundary $B$) are propagating almost freely to infinity. We call
`trapped' (or briefly T-modes) the $UP$-modes with $\nu < 4\mu _l
/\sqrt{27}$ which are mainly reflected by the potential barrier and
returned to the black hole's horizon.

Let $f_J^{UP}$ be a `visible' (either E- or T-) mode, i.e. a
mode which crosses $\Sigma_v$. Then the operation (\ref{5}) of averaging over
the states
corresponding to this mode gives the thermal density matrix
\begin{equation}
\hat{\rho}^H_J ={\normalsize Tr} ^{\mbox{\scriptsize{vis}}}_J
\hat{\rho}^{\mbox{\scriptsize{init}}}_J
=\rho ^0_J \exp \left[-2\pi \nu _J
\hat{\alpha}_J^{*SIDE} \hat{\alpha}_J^{SIDE}\right],  \label{18}
\end{equation}
$\nu _J$ being the frequency $\nu$ corresponding to the wave-packet
$J$.  The standard calculation shows that the contribution of this
mode $J$ to the entropy of a black hole is \cite{note0}
\begin{equation}
S_J =s(2\pi \nu _J), \hspace{1cm}
s(z)=\frac{z}{\exp z -1}-\ln [1-\exp (-z)]. \label{19}
\end{equation}
It is convenient to calculate the separate contributions
$S^E$ and $S^T$ of $E$ and $T$ modes and to write
\begin{equation}
S^H =S^E +S^T,
\end{equation}
where
\begin{equation}
S^{E,T} \approx \sum_{jn} \!^{'}\sum_{l=0}^{\infty} (2l+1)
\Theta [\pm (\nu _j^2 -16\mu _l^2 /27 )] s(2\pi \nu _j). \label{20}
\end{equation}
The sign $+$ in the step function $\Theta$ stands for $E$-modes, the sign
$-$ stands for $T$-modes, and the prime in $\sum_{jn} '$ means that the
summation is taken only over the modes which cross $\Sigma_v$.

In order to get $S^E$ we change the summation over $l$ by integration
over $\mu_l^2$ \cite{note3}
\begin{equation}
\sum_{l=0}^{\infty} (2l+1)\Theta (\nu _j^2 -16\mu _l^2 /27)
\approx \int_0^{27\nu_j ^2 /16}d\mu_l^2 =\frac{27}{16}\nu_j ^2. \label{21}
\end{equation}
The expression standing under the sum for $S^E$ does not depend
explicitly on $n$. Denote by $N_j^E$ the total number of different
values of the index $n$ for which a  mode with a given value of $j$
crosses  $\Sigma_v$. Then we have
\begin{equation}
S^E \approx (27/16)\sum_j N_j^E \nu_j^2 s(2\pi \nu_j). \label{22}
\end{equation}
In the DeWitt's approximation one can omit $V_l$ in Eq.(\ref{17}) and to get
that  $N_j^E  =(u_2 -u_1)\delta  /2\pi$, where  $u_{1,2}$ are  the values
of  the  retarded  time  $u=\eta  -\xi$  corresponding  to the boundary
points of the surface $\Sigma_v$.

The value $u_2$ is formally divergent.  The necessary cut-off arises
due to the quantum fluctuation of the horizon. Zero-point fluctuations
of the horizon result in its spreading so that due to the quantum noise
events happening closer than $x_{\lambda}=1+\lambda$ to the horizon cannot be
seen from outside.  One can show that
\begin{equation}
\lambda =\alpha l_{\mbox{\scriptsize{Pl}}}^2/r_g^2 ,\label{23}
\end{equation}
where $\alpha$ is a dimensionless coefficient
\cite{York:83,York:84,Beke:84}.    In   our context  it  means  that  we
are  to  consider as `visible'  only those particles  which cross $\Sigma
^{\lambda}_v$  which  is  the  part  of  $\Sigma_v$   lying outside
$x_{\lambda}$.
It should be stressed
that for black holes with $M\gg m_{\mbox{\scriptsize{Pl}}}$ (which we consider
here)
$\lambda \ll 1$.

By  using  this  cut-off  after  changing  the summation over $j$ by
integration ($\delta \sum_{j=0}^{\infty} =\int_0^{\infty}d\nu $) we get
\begin{equation}
S^E \approx
\frac{27}{32\pi} \Delta u \int_0^{\infty}d\nu \nu^2 s(2\pi \nu ),\label{24}
\end{equation}
where $\Delta u =q(x_0 )-\ln \lambda$  and the  function $q(x)$ is defined in
Eq.(\ref{16}).

Now  we  turn  to  the  calculation  of  the contribution $S^T$ of
$T$-modes to the black-hole's entropy. We write $S^T$ in the form
\begin{equation}
S^T \approx \int_0^{\infty}d\nu J(\nu )s(2\pi \nu ), \label{33}
\end{equation}
where
\begin{equation}
J(\nu_j) =\delta^{-1}\sum_{l=0}^{\infty}(2l+1)N_{jl}^T
\Theta (\frac{16}{27}\mu_l^2 -\nu_j^2 ),
\label{28}
\end{equation}
and  $N_{jl}^T$ is a number of $T$-modes  with given quantum numbers
$j$ and $l$ which cross $\Sigma^{\lambda}_v$.

Note  that due to  the exponential decrease of $s(z)$  the frequencies
$\nu \gg  1/2\pi$ do  not contribute   to   $S^T$.   For example, to
provide  the accuracy  $\sim 10^{-6}$ it is sufficient to consider the
frequencies $\nu <2.2$. Denote by $x(\nu )$ a turning point of a mode
of frequency $\nu$ ($V_l (x(\nu ))=\nu^2$).  For  large  $l$ one has
$x(\nu )-1\approx \nu ^2  /4\mu_l^2 \ll 1$. In the region $|x-1|\ll
1$ one  can  use the homogeneous-gravitational-field  (HGF)
approximation   and   put $W_l  =\mu_l^2$  in  Eq.(\ref{11})  and  $B=x=1$
in  the metric (\ref{7}). This approximate metric can be written in the
Rindler-like form
\begin{equation}
ds^2 =-\rho^2 d\eta^2 +d\rho^2 +d\omega^2, \label{28a}
\end{equation}
where $\rho =2(x-1)^{1/2}$.

In the HGF-approximation normalized solutions $F_{\nu l}^{UP}$
of Eq.(\ref{11}) which take the value $(4\pi \nu )^{-1/2}\exp (-i\nu u)$
at the horizon $H^-$ are
\begin{equation}
F_{\nu l}=A_{\nu l}\pi ^{-1}\sinh ^{1/2}(\pi \nu )
\exp (-i\nu \eta ) K_{i\nu}(\mu _l \rho ) ,
\label{25}
\end{equation}
where
\begin{eqnarray}
&&A_{\nu l} =-i \exp [-i (\nu \ln (\mu _l /2)-\phi _{\nu}], \nonumber\\
&&\exp (2i \phi _{\nu)}=\Gamma (1+i\nu )/\Gamma (1- i\nu ),
\label{26}
\end{eqnarray}
and $K_p (z)$ is a Macdonald function.

In order  to find  $N_{jl}^T$ we  note that  each of  the modes  which
crosses $\Sigma^{\lambda}_v$ crosses  also a space-like surface
$\Sigma^{\lambda}_t$ located
between  $p_2$  and  mirror-like  boundary  $B$  and  described by the
equation $\eta =$const . It is possible to show that
\begin{equation}
N_{jl}^T \approx 2\pi^{-2} \nu_j \delta \sinh (\pi \nu_j )
\int_{2\lambda ^{1/2}}^{\infty} d\rho \rho^{-1}K_{i\nu }^2 (\mu_l \rho ) .
\label{27}
\end{equation}
After substituting this relation  into Eq.(\ref{28}) and changing  summation
over $l$ by integration over $\mu_l^2$ we can rewrite this  expression
in the form
\begin{equation}
J(\nu ) \equiv J_a (\nu )
=\frac{\nu \sinh (\pi \nu )}{\pi ^2 \lambda }
\int_1^{\infty}\frac{dy}{y}\int_a^{\infty}
dz z K_{i\nu}^2 (zy) ,\label{29}
\end{equation}
where $a=(27\lambda )^{1/2}\nu /2$. For the frequencies $\nu$
which contribute to the black hole's entropy the parameter
$a$ is extremely small $a\ll 1$. We write
\begin{equation}
J(\nu ) =J_{a=0}(\nu )+\Delta J(\nu ),
\hspace{1cm}\Delta J(\nu ) =J_a (\nu )-J_{a=0}(\nu ) . \label{30}
\end{equation}
The integrals which enter $J_{a=0}(\nu )$ can be taken exactly and
we get
\begin{equation}
J_{a=0}(\nu ) =\frac{\nu ^2}{4\pi \lambda}, \label{31}
\end{equation}
while for $\Delta J(\nu )$ one can obtain the following
approximate expression
\begin{equation}
\Delta J(\nu ) \approx \frac{27\nu ^2}{32\pi}[\ln \lambda +
\ln (27\nu ^2 /16)-1 -\psi (1+i\nu )-\psi (1-i\nu )].
\label{32}
\end{equation}
Here $\psi (z)$ is the logarithmic derivative of $\Gamma$-function.

By adding this expression with the expression (\ref{24})
for $S^E$ we finally get
\begin{equation}
S^H =S^H_0 +\Delta S, \label{34}
\end{equation}
where
\begin{equation}
S^H_0 =\frac{1}{4\pi \lambda}\int_0^{\infty}d\nu \nu ^2 s(2\pi \nu )
=\frac{1}{360\lambda}, \label{35}
\end{equation}
\begin{equation}
\Delta S = \frac{27}{32\pi}\int_0^{\infty}d\nu \nu^2 Q(\nu )s(2\pi \nu ) ,
\label{36}
\end{equation}
and $Q(\nu )=x_0 -2 +\ln (27\nu ^2 /16)-\psi (1+i\nu )-\psi (1-i\nu )$.
Numerical calculations give $\Delta S\approx (9x_0 -23)\times
10^{-3}$.

One cannot expect to determine  the entropy with accuracy higher  than
it  is  allowed  by  uncertainties  related  with  its thermodynamical
fluctuations. That  is why  the term  $\Delta S$  in Eq.(\ref{34})  which is
much smaller than
 unity  can  be  neglected.  Thus one can identify the black hole's
entropy with $S^H_0$. By using Eq.(\ref{23}) we get
\begin{equation}
S^H_0 =\gamma \frac{A_H}{4\l_{\mbox{\scriptsize{Pl}}}^2},\hspace{1cm}
\gamma =\frac{1}{360\pi \alpha}, \label{37}
\end{equation}
where $A_H =4\pi r_g^2$ is the surface area of a black hole.
It is important to stress that $S^H_0$ does not depend on $r_0$ (the
raduis of the mirror-like boundary). It also does not depend on
the particular choice of the surface $\Sigma_v$ we introduced in order
to define the separation into 'visible' and 'invisible' modes.
For a stationary black hole the obtained result is evidently invariant
under the shift of the advanced time parameter $v$.
Instead of a null surface one can also use (without changing the result)
any space-like surface
crossing the horizon. The result (\ref{37}) reproduces the standard
expression for the black-hole's
entropy $A_H/4l_{\mbox{\scriptsize{Pl}}^2}$ for the value of the parameter
$\alpha =(360\pi )^{-1}$.

The calculation of the  coefficient $\alpha$ which enters  the expression
(\ref{23})  for  the   quantum  fluctuations  of   the  horizon  is  a
delicate  problem  which  requires  quantization  of  gravity.  The
following arguments allow one to estimate its value. By direct
measurements of black  hole's mass  $M$  during  time  interval  $\tau$
one  expects  a  measurement uncertainty   $\Delta   M   \approx  \xi
/\tau$  where  $\xi  \ge   1/2$ \cite{Beke:84}.  Spontaneous  quantum
emission (or
absorption)  of particles   by   a black  hole  results  in the
jumps $\sim (8\pi M)^{-1}$  of  the mass $M$. These jumps  do not  allow one
to
take   the  interval  $\tau$   as  long as  one   wishes  without generating
new additional uncertainties. The best accuracy in  a single measurement can
be
obtained if $\tau$ coincides with the time interval $t_1$ between two
subsequent events
of emission or absorption of quanta. The accuracy of defining of $M$ can be
improved
if instead of a single measurement one makes a sequence of measurements. Let
$N$ be
a number of independent measurements and $t_i$  $(i=1,\ldots N)$ be the time
intervals of these measurements (i.e. the time intervals between the emission
or absorption of quanta). The
probability   of  emission of  a  next  quantum at the    time
interval  ($t,t+dt$)    after    the    last  previous emission  is
$p(t)dt=\exp (-\bar{n}t)\bar{n}dt$,   where   $\bar{n}$ is the
average number of  quanta radiated in   a unit of   time.  An accuracy
of the black  hole's mass definition after large  number $N$ of
above described independent   measurements is \cite{Tayl:82}
\begin{equation}
\Delta _N M\approx N^{-1/2}(\overline{t^2})^{-1/2}
=(2N)^{-1/2}\bar{n}, \label{38}
\end{equation}
where $(\overline{t^2})=N^{-1}\sum_{i=1}^N t_i^2$.

In the above consideration it was inexplicitly assumed that the only origin
of the uncertainty of the measurement of the mass $M$ was connected with
the measurement procedure itself.  Now we note
that the quantum fluctuations of the horizon do not allow
to determine the mass $M$ of a black hole absolutely exactly even if there
would be no uncertainties in the measurement procedure itself.
The sequence of $N$ exact measurements of $M$ (in case if they
were possible) would give an accuracy
$\bar{\Delta}_N M =N^{-1/2}\delta M$ where $\delta M$
is the characteristic value of fluctuation of $M$ (its dispersion) connected
with quantum fluctuation of the horizon.
It is natural to assume that both accuracies (one ($\Delta _N M$) connected
with
the quantum uncertainty of the measurement procedure and the other
($\bar{\Delta}_N M$) connected
with quantum fluctuations of the horizon) are of the same order
of magnitude $\Delta _N M\approx \bar{\Delta}_N M$. Hence one can write
\begin{equation}
\delta M \approx 2^{-1/2}\xi \bar{n}. \label{39}
\end{equation}

The value of the average particles number rate of emission $\bar{n}$
for a scalar massless
field was numerically calculated by Simkins \cite{Simk:86} who found
\begin{equation}
\bar{n}=6.644\times 10^{-4}M^{-1} .\label{39a}
\end{equation}
(DeWitt's approximation gives very close result
$\bar{n}_{DeWitt}= 3^3 \zeta (3)/(2^9\pi ^4 M)
\approx 6.5\times 10^{-4}M^{-1}$.) The value of the parameter
$\alpha$ $(\delta M=\alpha /4M)$ corresponding
to the expression (\ref{39a}) is
\begin{equation}
\alpha \approx 1.88\times 10^{-3}\xi .\label{41}
\end{equation}
This result is quite close to the York's estimation of
the quantum fluctuations of the  event horizon based on the  deflection
of the  apparent horizon  from the  event horizon  for an  evaporating
black hole  \cite{York:84}.

By using Eq.(\ref{41})  one gets for  the coefficient $\gamma$  which enters
the
expression  (\ref{37})  $\gamma   \approx   0.47\xi   ^{-1}$.  For    the
oft-quoted  minimal value $\xi =1/2$ one has $\gamma  \approx 0.94$. This
estimation  is   in    a  good    accordance    with      the    exact
thermodynamical   value  of   the black-hole's entropy for which  $\gamma
=1$.

We make now some general remarks concerning the obtained result.   The
entropy  of   a  black   hole   is   of  pure   quantum  nature.   The
gravitational  field  of  a  black  hole continuously creates pairs of
particles. For alone static uncharged black hole one of the components
of a created pair is always located inside the horizon while the other
can escape  to  infinity  and  contribute  to  the  Hawking radiation.
For low frequencies the probability of escape is exponentially small so
that almost all of the components of such pairs created outside  the
horizon are reflected by  the potential barrier  and finally   fall
down  into the  black hole.  The existence of `invisible'  (hidden
inside the  horizon) modes results  in the  entropy  of  the  black
hole.  Eq.(\ref{2})  shows  that  only   those `invisible' components of
pairs contribute to the entropy of the black hole for  which the  other
component  is `visible'.  The life-time  of `visible' `trapped' modes
is of   order $\tau_{\nu l}\sim \kappa  ^{-1}\ln (l^2  /\nu ^2 )$. For
modes with  large $l\sim   \lambda  ^{-1/2}$ which give   the  main
contribution to the entropy    this `life-time' is  $\sim r_g  \ln (r_g
/\lambda  _{\mbox{\scriptsize{Pl}}}$). The  main   contribution to
the entropy of a black
hole is given by  `invisible' modes which are propagated in  the narrow
$\sim  \lambda _{\mbox{\scriptsize{Pl}}}$ shell region  near the horizon.

We estimated the contribution  to entropy of a  non-rotating uncharged
black hole by  a conformal massless  field. The generalization  to the
case  of  stationary (rotating and charged) black   holes  and
different physical fields  is straightforward. By using the above
arguments one can expect that  the number  of thermally exited trapped
modes  which  contribute  to  the entropy will be always proportional
to the surface area of a black hole.  If there are present more than one
fields each of them contributes to the entropy  additively.  On the other
hand  the average rate $\bar{n}$  of the emission of  the particles
grows as  the number  of fields $N$. That is why the parameter $\alpha$
which enters the expression (\ref{37}) for the entropy of a black hole
and which characterizes the quantum
fluctuations of the horizon also grow as $N$.  For a large number $N$
of fields these two  effects cancel each  other so that  the entropy
does not depend on the number of fields.

In conclusion  we return  to the  gedanken experiment  with a wormhole
\cite{FrNo:93} discussed at the beginning.  The entropy of a black hole
(identified with  its surface area)  decreases when  one of  the wormhole's
mouths is falling inside a black hole and returns to its initial value
after another mouth crosses the horizon or the wormhole is destroyed.
The proposed interpretation of `invisible' modes  as dynamical  degrees
of freedom of  a black hole allows one to  understand this `mysterious'
behavior of the entropy. In priciple byy using a wormhole one can obtain an
additional information concerning the quantum  states  inside  the  black
hole.  The  total number  of the originally `invisible' modes
propagating near the gravitational radius which become `visible' via
the wormhole is proportional to the  change of the black hole surface
area. The decrease of number of  `invisible' states results in the
decrease of the entropy of the black hole.  In principle  by using  a
wormhole  one can  change the states of some of the originally
`invisible' modes without changing external parameters of the black hole
\cite{note4}. This
possibility is lost after the second mouth of the wormhole crosses the
horizon or the wormhole is destroyed. The `visible' components  of such
excited modes continue living outside the horizon only short time
comparable with their `life-time'. After this they fall down into  the
black  hole  and  the corresponding  pair  does  not contribute to the
entropy of the black hole. The system `forgets' an intervention  (if
only  it did  not  change  the black hole parameters) and  the entropy
of the  black hole  returns to  its initial value.

\vspace{.5cm}
{\em Note added:} -- After this paper was submitted for publication the
paper by Srednicki \cite{Sred:93} appeared. For the massless scalar
field in a flat spacetime he showed that the entropy
arising after tracing the degrees of freedom of the field in the vacuum
state residing inside a sphere is proportional to the area of this sphere.
The analogous result was obtained earlier in Ref.\cite{BoKoLeSo:86}.
In the present paper we have shown that the main contribution to the
black-hole's entropy is given by modes, propagating near its surface,
and hence the entropy can also be considered as the 'surface effect'.
Despite the formal similarity of the results of the flat spacetime
and of the black hole calculations there is a big difference between them.
The black hole horizon is a null surface.
Its geometry differs from the geometry of a timelike surface in a flat
spacetime.
The physical meaning and the mathematical description of modes which
contribute
to the entropy are quite different in both cases.
That is why it is not clear how far interesting arguments of
Refs.\cite{Sred:93,BoKoLeSo:86}
based on the flat spacetime calculations can be
directly applied to black holes.

\vspace{1cm}

{\bf Acknowledgements:} The authors are grateful to Werner Israel for
stimulating discussions. They also thank Don Page for helpful comments
and his help in finding the Ref.\cite{Simk:86}.
This work was supported in part by
the Danish Natural Sciences Research Council, grants N11-9640-1 and
11-9524-1SE and also by the Natural Sciences and Engineering Research
of Canada and by Deutsche Forschungsgemeinschaft (V.F.).


\begin{thebibliography}{99}
\bibitem{note0} Strictly speaking $s(z) =(1+n)\ln (1+n)-n\ln (n)$,
where $n=\Gamma (\exp z -1)^{-1}$, and $\Gamma$ is the probability
for the wave packet to reach $\Sigma_v$. For $\Gamma =0$ $s(z)=0$,
while for $\Gamma =1$ one gets the expression (\ref{19}). The approximation
we use is analogous to the one proposed by DeWitt
(Phys.Reports {\bf 19C}, 295 (1975)). We put $\Gamma =1$ for the modes
which reach $\Sigma_v$ and $\Gamma =0$ for modes which do not reach it.
We shall see that the main contribution to the entropy of a black hole
is given by the modes propagating in the close vicinity of the horizon.
Because of the very high blue-shift effect for these modes one can
use geometric optics approximation and verify that the approximation
we use does give the correct result.
\bibitem{note4} The $SIDE$-modes are propagating in the region lying to
the future with respect to the horizon, while the $UP$-modes are (at least
partly)
propagating in the region lying to the past with respect to the horizon.
Due to the evident causal asymmetry between $UP$ and $SIDE$-modes
it is not quite clear whether one can also change the states of $T$-modes
without any changing of the black hole parameters and whether
these $T$-modes might give an additional contribution to the entropy of
the black hole. In any case because of the presence of quantum correlations
between $UP$-modes  and $SIDE$-modes, the total entropy $S$ of the system
cannot
be obtained simply by adding $S^H$ and the entropy calculated for the density
matrix (\ref{1}).
\end{thebibliography}
\end{document}